\documentstyle[prl,aps]{revtex}
\input epsf

\begin{document} \draft 
\twocolumn 
\title{Exploring a quantum degenerate gas of fermionic atoms} 
\author{B. DeMarco and D. S. Jin\cite{adr1}} 
\address{JILA, \\
National Institute of Standards and Technology and University of
Colorado,
\\ and
\\ Physics Department, University of Colorado, Boulder, CO
80309-0440} 
\date{\today} 
\maketitle 
\begin{abstract} 

We predict novel phenomena in the behavior of an ultracold,
trapped gas of fermionic atoms.  We find that quantum statistics
radically changes the collisional properties, spatial profile, and
off-resonant light scattering properties of the atomic fermion system,
and we suggest how these effects can be observed.

\end{abstract} 

\narrowtext

The experimental ability to cool a dilute gas of atoms to temperatures
well below one microKelvin has introduced a new area of research,
highlighted by the observation and study of Bose-Einstein condensates
(BEC) \cite{BEC1,BEC2,BEC3}. Interparticle interactions in these quantum
degenerate gases are weak and therefore amenable to theoretical treatment; 
in addition, the interaction strength can be altered experimentally
\cite{interactions,Feshbach}. A quantum degenerate trapped gas of
fermionic atoms, i.e., atoms with an odd total number of constituent
fermions, is relatively unexplored theoretically as well as
experimentally. An intriguing possibility is that at
sufficiently low temperature the atoms will develop pairwise correlations,
analogous to Cooper pairing of electrons, and undergo a phase transition
to a superfluid-like state.  Theoretical studies have concluded that this
new fermionic condensate will only occur at very low temperatures and will
be difficult to observe experimentally \cite{FDBEC}.  We explore a
trapped gas of fermionic atoms that is quantum degenerate but
uncondensed, a regime that is interesting in its own right. 

Experimental techniques similar to those used in trapping and cooling
bosonic atoms such as $^{87}$Rb, $^{23}$Na, or $^7$Li can be employed to
trap and cool fermionic atoms, with possible experimental candidates
including $^{40}$K, $^6$Li, $^{2}$H, and metastable $^3$He. We focus on
$^{40}$K, which has a large number of trappable spin states, as an example
for the effects discused herein.  We consider $10^6$ atoms in a
cylindrically symmetric harmonic trap with a radial trap frequency
$\omega_r=2\pi\times 400$Hz and an axial frequency $\lambda\omega_r$ where
$\lambda = 0.1$.  The characteristic temperature for quantum degeneracy,
the Fermi temperature, is given by $T_F={\hbar\omega_r \over k_B}(6\lambda
N)^{1/3}=1.6\mu$K \cite{Fermitemp}. (Note that the criteria for quantum
degeneracy for fermions is approximately the same as for bosons, since the
BEC phase transition occurs at $T_c={\hbar\omega_r \over k_B}(\lambda
N/1.202)^{1/3}$).  While there is no phase transition for fermions at
$T_F$, the behavior of the system changes below $T_F$ where quantum
statistics becomes important. 

We first examine collisional properties of the trapped, fermion gas.  In
the usual BEC experiment, the atomic sample is cooled optically then
loaded into a magnetic trap.  To avoid losses due to spin-exchange
collisions, typically only atoms in a single spin state are loaded into
the trap.  Cooling then proceeds by forced evaporation, i.e.,  removal of
the most energetic atoms and rethermalization of the gas via elastic
collisions. For bosonic atoms at these low temperatures the elastic
collisions are predominately binary and {\it s}-wave in character. For
fermionic atoms the dominant contribution must be {\it p}-wave since
Fermi-Dirac (FD) statistics prohibits {\it s}-wave collisions between
identical particles.  However, the {\it p}-wave collision cross-section is
suppressed at low temperature due to the centrifugal barrier. The
threshold collision energy $E_{th}(l)$ for a given partial wave $l$ to
contribute to the scattering can be approximated by the height of the
centrifugal barrier, \begin{eqnarray} E_{th}(l)={\hbar^2 l(l+1) \over
2mb^2}-{C_6 \over b^6}, \;\;b^2=({6C_6m \over \hbar^2 l(l+1)})^{1/2}.
\end{eqnarray} Using the $C_6$ coefficient for K-K collisions \cite{csix}
gives a {\it p}-wave threshold energy $E_{th}(l=1)=100 \mu$K. Thus, below
100 $\mu$K, the elastic collision rate in a gas of identical $^{40}$K
atoms will rapidly vanish, unless there exists an accidental {\it p}-wave
resonance. 

Sympathetic cooling \cite{Myatt} can be used to circumvent this obstacle
to evaporative cooling.  If bosonic atoms as well as fermionic atoms are
loaded into a magnetic trap, the bosonic atoms can be cooled evaporatively
and the fermionic atoms will be cooled ``sympathetically"  via ({\it
s}-wave)  elastic collisions with the bosons.  A second alternative is to
load two species of fermionic atoms into the trap, allowing {\it
s}-wave elastic collisions between non-identical fermions to rethermalize
the gas during evaporative cooling. For the remainder of this paper, we 
assume one can successfully cool a gas of fermionic atoms to the
quantum degenerate regime and that the background gas used for sympathetic
cooling can be removed, without significantly disturbing
the fermionic gas.

The quantum statistical suppression of {\it s}-wave interactions between
identical fermions makes an ultracold, trapped gas of fermionic atoms an
excellent realization of an ideal gas. This could be seen experimentally
with a measurement of the elastic rethermalization rate \cite{elastic} of
a single component compared to a two-component (bosonic and fermionic
atoms, or two species of fermionic atoms) gas.  Alternatively, a study of
shape excitations of the trapped atom cloud \cite{excitations} will reveal
this effect. For a gas of identical fermions in a harmonic trap, shape
oscillations will occur at exact multiples of the characteristic trap
frequency and will damp at a rate determined by the elastic collision rate
in the cloud. At temperatures below the threshold for {\it p}-wave
collisions ($\approx 100\mu$K), which is still well above the temperature
of quantum degeneracy, the ratio of the {\it p}-wave to {\it s}-wave
elastic cross-section $\sigma_p/\sigma_s$ scales as $(ka)^4$ where $k$ is
the atomic wave vector and $a$ is a characteristic range of the potential.
For example, a collision energy of 100$\mu$K and a range $a=50$ Bohr radii
gives $\sigma_p/\sigma_s\approx 10^{-3}$.  The excitation lifetimes at low
T, therefore, will be limited by the lifetime of atoms in the trap
(typically 100's of seconds).  For comparison, in a two-component fermion
cloud these excitations will have shorter lifetimes (a reasonable estimate
is $\approx 100$ ms, the lifetime seen for shape oscillations in
non-condensed, bosonic atom clouds \cite{excitations}) and at
sufficiently high densities should exhibit a small frequency shift
\cite{Stringariex} due to the {\it s}-wave interactions.  Interestingly
enough, anomalously low damping rates could be observed in a dilute gas of
fermions before they are observed in atomic BEC (due to different
physics).

Quantum statistics also changes the collisional properties of
multi-component fermionic atom clouds.  Atoms in two (or more) spin states
could either be loaded initially into a magnetic trap or created by an rf
or microwave pulse applied to a cold single-component cloud.  At low
temperatures ($T<100\mu$K) quantum statistics modifies the inelastic
collisional loss rates by prohibiting {\it s}-wave spin-exchange
collisions involving an initial or final state consisting of identical
fermions.  For example, $^{40}$K atoms in the two spin states
$|F=9/2,m_F=9/2 \rangle$ and $|9/2, 7/2 \rangle$ are stable against spin
exchange at low temperatures (note the inverted ground-state structure
shown in Fig. 1).  Furthermore, at quantum degenerate temperatures
otherwise allowed spin-exchange collisions can be suppressed through final
state occupation.  For example, an ensemble of $^{40}$K atoms in the three
spin states $|9/2,9/2 \rangle$, $|9/2,7/2 \rangle$, and $|9/2,5/2 \rangle$
is subject to the following spin-exchange collision:  $|9/2,5/2 \rangle +
|9/2,7/2 \rangle \rightarrow |9/2,3/2 \rangle + |9/2,9/2 \rangle$.  This
process will limit the lifetime of the $|9/2,5/2\rangle$ atoms in the
trap.  However, for $T/T_F\ll 1$ the Pauli exclusion principle will
suppress this $m_F$ changing collision since the final state, $|9/2,9/2
\rangle$ with an appropriate energy and momentum conserving external
state, is likely to be occupied.  This suppression of spin-exchange could
be observed as an increased lifetime for the $|9/2,5/2\rangle$ atoms in
the three-component cloud at quantum degenerate temperatures. 

\epsfxsize=3 truein 
\epsfbox{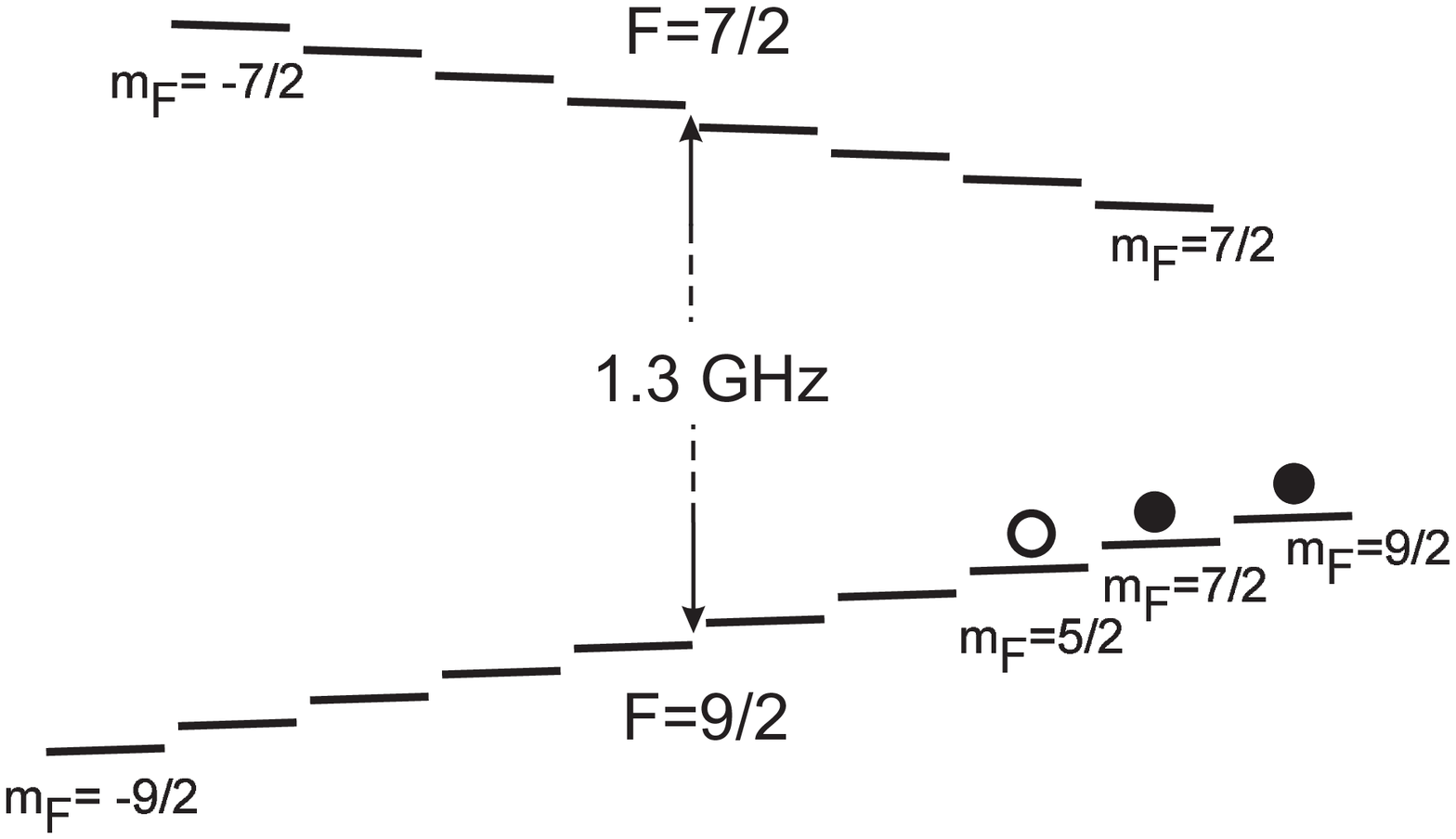} 
\begin{figure} \caption{Schematic of the ground-state hyperfine levels
of $^{40}$K (shown with exaggerated Zeeman splittings). 
\label{fig1}}
\end{figure}

We now consider the interaction of atoms with light, which is a powerful
tool for probing trapped atomic gases and has been studied theoretically
for ultracold bosons, and to a lesser extent, fermions
\cite{light1,light2,fatline,blocking,Juha}. One effect of quantum
statistics on light scattering, which has been essential for studying BEC
with optical imaging techniques, is simply that the scattered light
reflects the spatial profile of the atom cloud, which in a harmonic trap
is directly related to the quantum statistical occupation of trap energy
levels. For example, light scattering images of trapped, bosonic atoms
reveal deviations from the classical gaussian spatial profile even at
temperatures slightly above the BEC phase transition
\cite{BEC3,Boseprofile}.  The analog of this effect for fermions provides
a straight-forward measure of the quantum statistics, which is useful to
quantify for comparison to other probes of quantum degeneracy. 

A trapped fermion cloud will have a larger spatial extent and a lower peak
density relative to Maxwell-Boltzman (MB) particles (see Fig. 2 inset). 
This effect will be small unless T is well below T$_F$
\cite{Fermitemp,Fermiprofile,Rokhsar}, but may be revealed in careful
studies of the cloud profile as a function of temperature.  For example
one could observe the momentum distribution by imaging the atoms after a
period of free expansion from the trap\cite{BEC1}. The expected
two-dimensional expanded cloud image can be calculated from the
semiclassical momentum distribution for fermions in a harmonic trap
\cite{Rokhsar}, integrated through in one dimension.  For sufficiently
long expansion times the initial spatial distribution can be ignored.
Interactions can also be ignored for a single-component fermion cloud for
the reasons discussed previously. A simple measure of the extent to which
FD statistics makes the expanded cloud image non-gaussian can be extracted
by comparing a gaussian surface fit to the entire image and a fit to only
the outer wings of the cloud image.  Fitting the full cloud image reveals
the enhanced width due to FD statistics (larger average energy) while a
fit to the outer edges of the cloud gives a width that more closely
reflects the cloud temperature.  In Fig. 2 we plot the
ratio of the gaussian widths from these two fits, as a function of reduced
temperature $T/T_F$.
	
\epsfxsize=3 truein 
\epsfbox{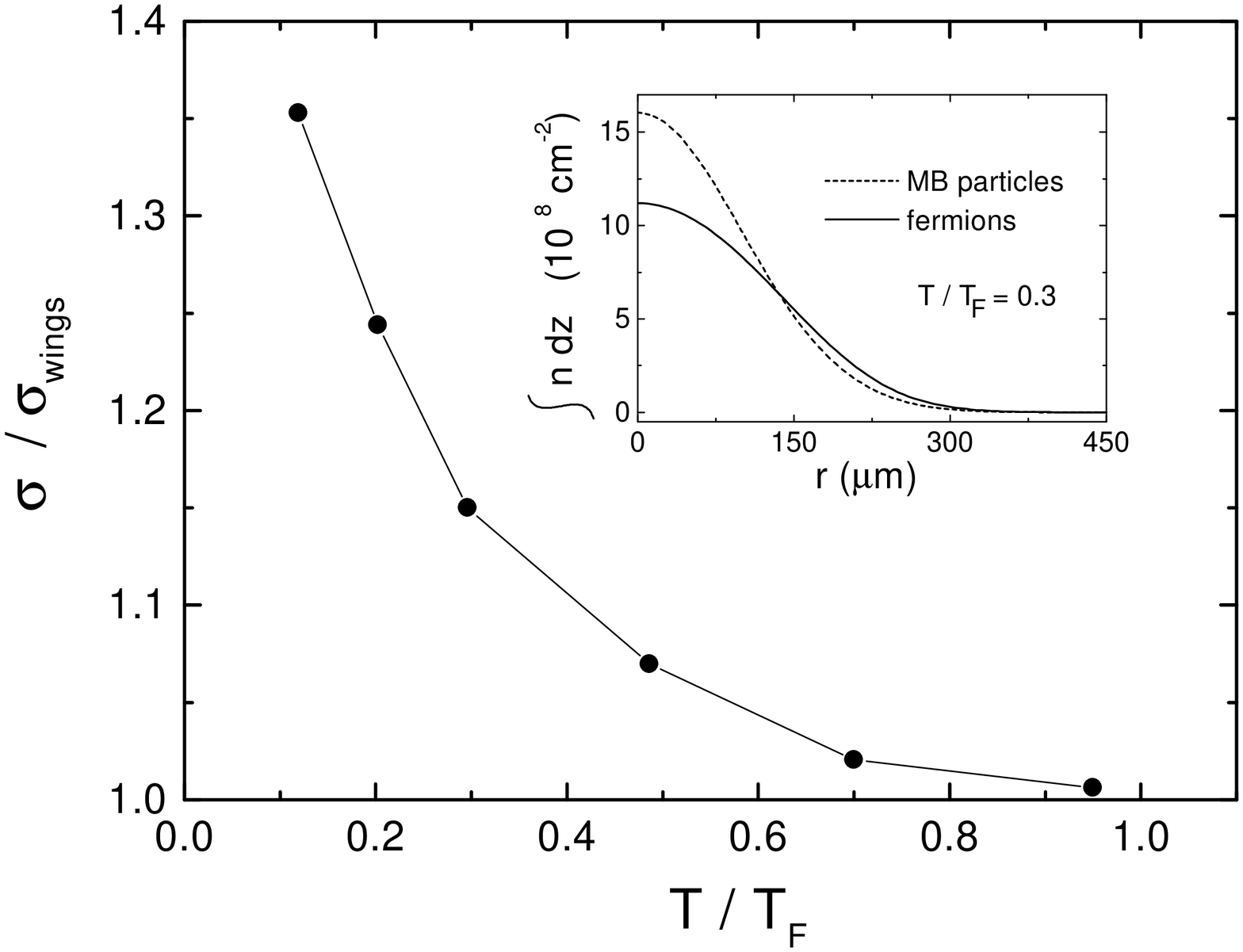} 
\begin{figure} \caption{Ratio of the widths derived from gaussian surface
fits to the full expanded fermion cloud image and to only the outer edges
of the cloud $\sigma/\sigma_{wings}$ vs. $T/T_F$.  The wings were chosen
to include approximately one fifth of the total number of atoms.  The
inset shows traces through the cloud image for FD and MB particles,
calculated for$^{40}$K, $T/T_F=0.3$, and a 10 $\mu s$ expansion.
\label{fig2}}
\end{figure}

A second effect of atom quantum statistics on light scattering comes from
an enhanced or decreased probability for scattering atoms into occupied
trap levels.  Assuming a two-level atomic system, an atom that scatters a
photon receives recoil momentum that projects it onto new harmonic trap
levels, which may or may not be already occupied. (Experimentally a
two-level system may be approximated by driving the atomic cycling
transition with circularly polarized light whose quantization axis
coincides with the direction of the trap magnetic field.) For bosons an
enhanced probability of scattering into occupied final states is predicted
to give an extremely broad line width \cite{fatline}.  For fermions, the
Pauli exclusion principle does not allow scattering into occupied final
states resulting in a blocking effect \cite{blocking,Juha,Zoller} which
implies a narrowing of the linewidth. 

We explore this effect by calculating the scattering rate of weak,
off-resonant, monochromatic light from an optically thin cloud of
fermionic atoms.  This subject has also been treated by Javanainen and
Ruostekoski \cite{Juha} who look at the spectrum of light scattered from a
spatially homogeneous gas.  We focus here instead on the angular
distribution of scattered light, and treat a harmonically confined gas. We
start with the spectral density function (Eq.  47 in \cite{Juha}) 
describing the scattered light intensity at ${\bf r}$ with frequency
$\omega_L+\omega$, where $\omega_L$ is the frequency of the probe laser,
\begin{eqnarray} {\bf S}({\bf r},\omega)= {1 \over 2\pi}I(r){R^2 \over
\delta^2}{\bf M}( \hat{e},\hat{n})  \int dt\int d^3r_1 \nonumber\\
\times\int d^3r_2 \exp^{i\omega t+i{\bf \Delta k}\cdot({\bf r}_1-{\bf
r}_2)} \nonumber\\ \;\;\times\langle\psi^\dagger({\bf r}_1,0)\psi ({\bf
r}_1,0) \psi^\dagger({\bf r}_2,t)\psi({\bf r}_2,t)\rangle, \end{eqnarray}
where $\hbar{\bf \Delta k}$ is the recoil momentum of the atom, ${1 \over
2\pi}{R^2 \over \delta^2}I(r){\bf M}(\hat{e},\hat{n})$ is the scattered
intensity from a single free atom, and $\psi$ denotes the atomic
ground-state field operator.  In Eqn. 2, the excited-state field operators
have been replaced by an adiabatic solution in terms of the ground-state
field operators \cite{Juha}.  We then use the finite temperature version
of Wick's theorem \cite{Fetter} to evaluate the expectation value of the
four field operators, taken with respect to the atoms' many-body
wave function. We use the unperturbed Green's function for fermionic atoms
confined in a harmonic trap, and ignore coherent scattering which is
relevant only for very small scattering angles and does not depend on the
atom quantum statistics. Taking the ratio of the scattered intensity for
FD compared to MB particles gives \begin{eqnarray} { {\bf S}_{FD} \over
{\bf S}_{MB}}(\theta,\phi,\omega)=
 {1 \over N}\sum_{{\bf i}} \sum_{{\bf n}} [1-f(E_n,T)] f(E_i,T) 
\nonumber\\ \times|\int d^3p \phi^*_{\bf n}({\bf p}+\hbar{\bf \Delta
k})\phi_{\bf i}({\bf p})|^2\delta(E_n-E_i-\hbar\omega). \end{eqnarray}
Here ${\bf i}=(i_x, i_y, i_z)$
and ${\bf n}=(n_x, n_y, n_z)$ denote the initial and final harmonic trap
levels of the atom, $f(E,t)={1 \over z\exp^{E \over
k_BT}+1}$ is the FD function, and $z$ is the fugacity.  The single
particle, momentum-space wave function for a harmonically confined atom is
denoted by $\phi_{\bf n}({\bf p})$, and the integral above has an analytic
form \cite{integral}.  Eqn. 3 can also be understood from Fermi's Golden
Rule, where one integrates over all possible final states ${\bf n}$ and
averages over initial states ${\bf i}$.  Experimentally one would detect
the scattered photons, with the detector position defining the scattering
angle ($\theta,\phi$) relative to the incident probe beam (along
$\hat{x}$) and conservation of momentum determining the atom recoil. For
$\lambda_L=767$ nm light at the $^{40}$K cycling transition, the recoil
energy $E_{recoil}={h^2 \over 2m\lambda_L^2}$ is $21\hbar\omega_r$.  The
ratio $E_{recoil}/E_F$, an important parameter in characterizing the light
scattering properties, is then 0.25 for our calculations. 

Figure 3 shows the angular dependence of the scattered light intensity,
integrated over all scattered photon frequencies, for fermions relative to
that for MB particles. (For the temperatures we consider the scattering is
essentially isotropic in $\phi$, reflecting the isotropy of the momentum
distribution at thermal equilibrium.) The blocking effect of FD statistics
appears as a suppression of the photon scattering rate, predominantly at
small angles where the small recoil momentum means that the final states
are mostly occupied.  This effect could be studied by
monitoring the scattering rate at small angle as a function of
temperature.  Alternatively, one could measure the ratio of the scattered
light intensity at a forward angle to that at a backward angle, revealing
the enhanced backward scattering at low reduced temperatures
\cite{zeroTnote}.  In addition to modifying the angular dependence of the
scattered light, FD statistics also suppresses the total scattering rate
(see Fig. 3).  At $T/T_F=0.24$, for instance, the lifetime of the atomic
excited state grows to 1.7 times the natural atomic transition lifetime.
This effect is reminiscent of enhanced atomic excited state lifetimes seen
in cavity experiments \cite{cavity}, although in our case the lifetime
grows because of a suppression of atom, rather than photon, final states.
The spectrum of scattered light, while more challenging to
measure, can be calculated straight-forwardly from Eqn. 3. 
The blocking effect of FD statistics shifts the spectrum to lower
frequency as shown in the inset to Fig. 3.     

\epsfxsize=3 truein 
\epsfbox{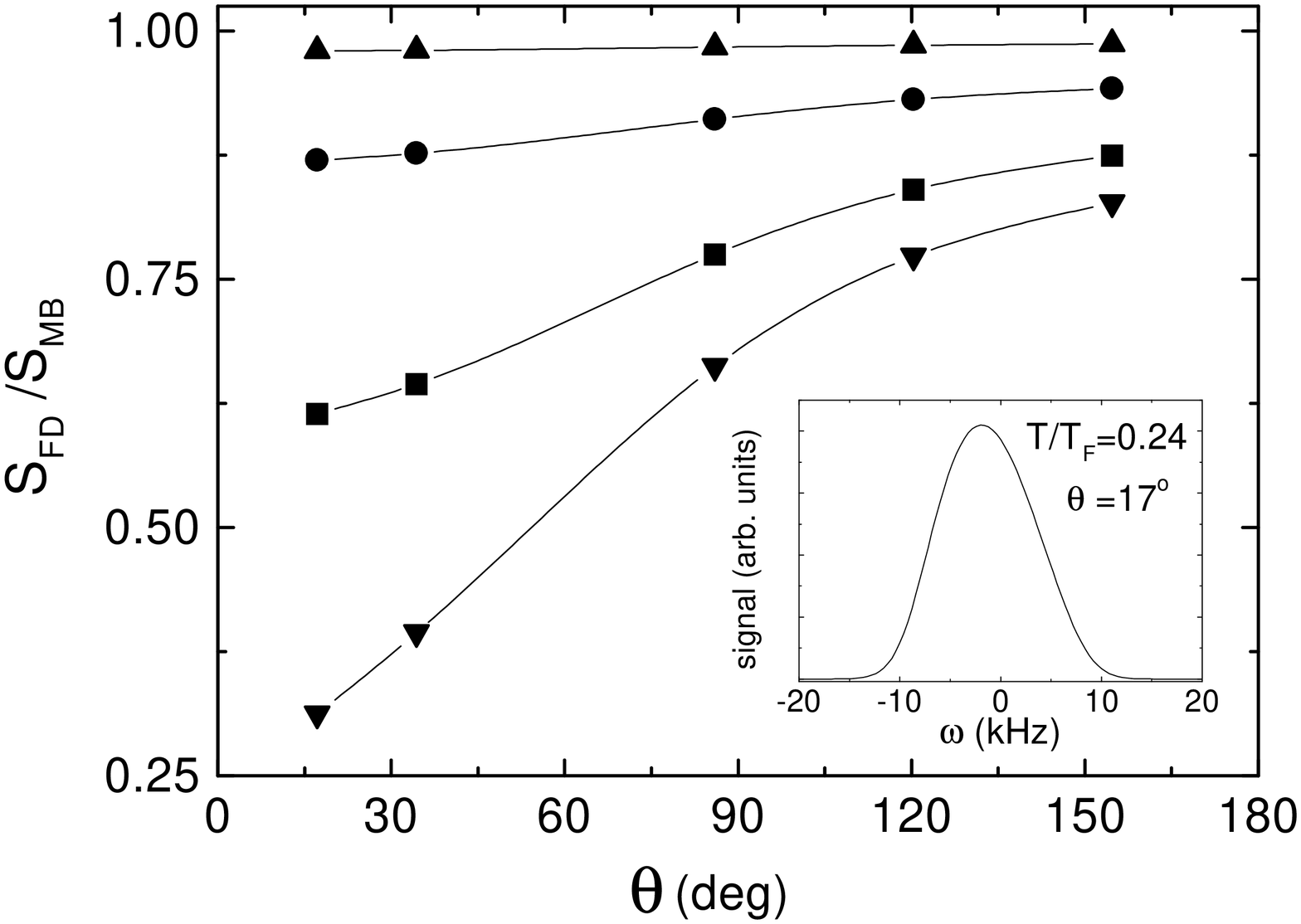} 
\begin{figure} \caption{Angular
dependence of scattered light intensity for trapped, fermionic atoms,
relative to that for classical particles, shown for
$E_{recoil}=21\hbar\omega_r$, $E_F=84\hbar\omega_r$, and $T/T_F$=1.0 (up
triangles), 0.47 (circles), 0.24 (squares), and 0.10 (down triangles). The
error on the calculated points, due to truncating the summation in Eqn. 3,
is less than 0.5 \%.  The inset shows the spectrum of the scattered light
for $T/T_F=0.24$ and $\theta=17^{\circ}$.  
\label{fig3}} \end{figure}

The large impact of FD quantum statistics on light scattering can
be used to probe quantum degeneracy. A large blocking effect requires both
small $T/T_F$ (see Fig. 3) and $E_{recoil}$ less than or comparable to
$E_F$.  This second condition implies that the experiment is best
performed with large $N$ and a tightly confining trap. Another way to
enhance the observed effect is to use a focused probe beam with a waist
smaller than the cloud size, thus interrogating atoms in lower harmonic
oscillator levels preferentially. 

In conclusion, we have examined the emergence of quantum statistical
effects on the behavior of a trapped gas of fermionic atoms.
Collisional properties as well as light scattering exhibit striking
changes, such as the suppression of radiative and non-radiative decay
processes, and provide convenient probes of quantum degeneracy. 

This work is supported by the National Institute of Standards and
Technology and the National Science Foundation. The authors would
like to express their appreciation for useful discussions with
M. Holland, M. Chiofalo, C. Wieman, E. Cornell, and J. Bohn.



\begin{references}
\bibitem{zeroTnote} In the limit of $T=0$, forward scattering
will actually become the dominant process as atoms are forced to
scatter into their initial harmonic oscillator states.  Our
calculation, however, does not include any effect of the probe beam on the
occupation of trap energy levels.
\end{references}
\end{document}